\documentstyle[twocolumn,prc,aps]{revtex}
\input epsf
\begin{document}
\twocolumn[\hsize\textwidth\columnwidth\hsize\csname @twocolumnfalse\endcsname
\title{Transverse Momentum of $\psi$ and Dimuon Production 
in Pb+Pb Collisions}
\author{Sean Gavin$^{a,b}$ and Ramona Vogt$^{c}$\cite{rv}}
\address{
$^a$Physics Department, Brookhaven National Laboratory, Upton, New York, 
11973\\
$^b$Physics Department, Columbia University, New York, New York, 10027\\
$^c$Institute for Nuclear Theory, University of Washington,
Seattle, Washington, 98195}
\date{\today}
\maketitle
\begin{abstract}
CERN collaboration NA50 has measured charmonium and Drell--Yan dimuon
production in Pb+Pb collisions.  Parton scattering broadens the
transverse momentum, $p_{{}_{T}}$, distributions for these processes.
We predict that $\langle p_{{}_{T}}^{2}\rangle$ will flatten in Pb+Pb
collisions as a function of the neutral transverse energy of hadrons,
$E_{T}$, in contrast to the almost--linear rise seen in
S+U~$\rightarrow \psi +X$.  If seen, such a flattening will support
hadronic explanations of charmonium suppression.

\vspace{0.1in}
{\bf preprint: CU-TP-791, DOE/ER/40561-292-INT96-21-01}
\end{abstract}

\pacs{CU-TP-791}

]

\begin{narrowtext}

The NA50 collaboration at the SPS has reported a suppression of $\psi$
production in Pb+Pb collisions relative to Drell--Yan dimuon production 
as the neutral transverse energy of hadrons, $E_{T}$, is
increased \cite{na50}.  They further presented a striking `threshold
effect' by comparing the data to S+U results as a function of a
calculated quantity, the mean path length of the $\psi$ through
nuclear matter, $L$.  The NA50 comparison is shown in fig.~1; the open
symbols are from ref.~\cite{na50}.  In this note, we demonstrate that
the relation between $L$ and the measured $E_{T}$ is model dependent.
However, we point out  that measurements of the centrality dependence of the
transverse momenta of Drell--Yan dimuons essentially provide an
experimental determination of $L$.  We argue that such a determination
can provide vital evidence for -- or against -- the threshold
behavior, which has been linked to the onset of quark--gluon plasma
formation in the Pb system \cite{na50,bo}.

To see how the path length affects the `threshold' interpretation of
the NA50 comparison, we replot the Pb+Pb data using $L(E_T)$
calculated with eq.~(\ref{eq:let}) and the realistic nuclear densities
\cite{jvv} that were employed in \cite{gv2}.  We see that the
appearance of the data is very sensitive to small changes in the
definition of $L$.  With the realistic $L$, one no longer gets the
impression that the Pb+Pb data ``departs from a universal curve.''
This departure has been cited as evidence \cite{na50} that Pb
collisions cross a threshold and form quark--gluon plasma.

In ref.~\cite{gv2}, we observed that the Pb+Pb data are in good accord
with predictions of charmonium suppression using a hadronic comover
model \cite{gv,gstv}.  While these predictions are supported by
cascade calculations \cite{cascade}, the plasma explanation cannot be
excluded.  We attributed the behavior of fig.~1 to a geometric
saturation that occurs in the symmetric Pb+Pb system, but not in
asymmetric S+U collisions; we discuss this below.  We see in fig.~1
that the saturation phenomena, though softened, does not vanish with
our calculated $L(E_{T})$.  Saturation -- the flattening of $L(E_{T})$
-- indeed occurs in our scenario, but only as a consequence of
geometry.
  
Important information on the suppression mechanism can be extracted
from the nuclear dependence of the $\psi$'s transverse momentum,
$p_{{}_{T}}$.  The hadronic suppression in ref.~\cite{gv2,gv} does not
modify the transverse momentum dependence of $\psi$ production
appreciably \cite{ggj}.  To account for the nuclear modification
measured in $p$A and S+U collisions, initial--state parton scattering
was introduced \cite{gg,pt}.  Bodwin, Brodsky and Lepage and,
independently, Michael and Wilk had predicted that initial-- and
final--state parton scattering can modify the momentum distributions
of hard interactions in nuclear collisions \cite{brod}.  Evidence for
such scattering has been seen in a variety of hadron--nucleus
experiments \cite{pa,pat,coca}.
\begin{figure} 
\vskip -1.2in
\epsfxsize=3.5in
\centerline{\epsffile{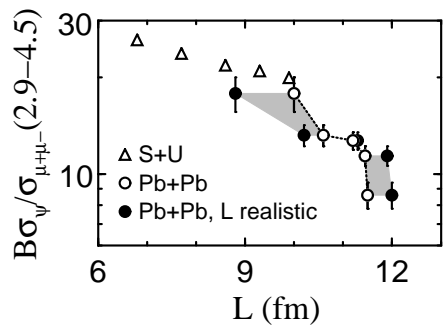}}
\vskip -1.6in
\caption[]{Comparison of NA50 Pb+Pb (open circles) and NA38 S+U (open
triangles) data as in \cite{na50}.  Data replotted with a realistic
$L(E_T)$ from eqs.~(\ref{eq:let},\ref{eq:nuet}) appear different. The
gray area represents the uncertainty in this presentation of the data
due to $L(E_{T})$.}
\end{figure}

Initial--state scattering \cite{brod} is elastic at the parton level
and broadens the $p_{{}_{T}}$ distributions of charmonium and
Drell--Yan production without affecting the $p_{{}_{T}}$--integrated
yields \cite{gg,pt}.  Partons essentially undergo a random walk in
momentum space, so that $\langle p_{{}_{T}}^{2}\rangle$ grows linearly
with $L$.  Initial--state scattering provides the only nuclear effect
proposed so far that can modify the $p_{{}_{T}}$ distribution in the
Drell--Yan process in a manner consistent with data \cite{pa,pat}.
While other effects such as comover scattering can alter the
$p_{{}_{T}}$ distribution in charmonium production, we find in
practice that the contributions to $\langle p_{{}_{T}}^{2}\rangle$ are
small if the comovers are assumed to be hadrons (see \cite{gg} and the
discussion below).  Measurements of $\langle p_{{}_{T}}^{2}\rangle$ --
particularly for dimuons -- can therefore provide {\it experimental}
information on $L$.

If $\langle p_{{}_{T}}^{2}\rangle$ increases with $L$ as in hadronic
models of charmonium suppression incorporating initial--state
scattering, then the geometric saturation in Pb collisions implies a
flattening of $\langle p_{{}_{T}}^{2}\rangle$.  In contrast, if the
threshold interpretation \cite{na50} is correct and the Pb results
are the consequence of a new contribution to $\psi$ suppression from
quark--gluon plasma, then one would not expect $\langle
p_{{}_{T}}^{2}\rangle$ to flatten with $E_{T}$.  If observed, a
flattening would lend strong support to the comover explanation.
Alternatively, if no flattening is observed, then our hadronic model
would be excluded.  Plasma suppression is currently thought to be
strongly $p_{{}_{T}}$ dependent \cite{plasma}, although models with
a weaker dependence (and without thresholds) have been presented
\cite{matsui}.

Plasma issues aside, measurements of $\langle p_{{}_{T}}^{2}\rangle$
can help remove the model dependence of fig.~1 \cite{na50}.  Gerschel
and H{\"u}fner \cite{gh} pointed out that a plot similar to fig.~1 but
containing only measured quantities can be constructed by replacing
$L$ with the measured $\langle p_{{}_{T}}^{2}\rangle$ of the $\psi$.
Measurements of the $\langle p_{{}_{T}}^{2}\rangle$ of the Drell--Yan
process, though more difficult, can provide a clearer determination of
$L$.  

In this paper we update the description of initial--state scattering
from ref.~\cite{gg} using more recent information from FNAL $p$A experiment
E772 \cite{pat}.  We then derive an expression for the path length
$L$, and study its behavior as a function of $E_T$ for S+U and Pb+Pb
collisions.  Finally, we combine these results to predict the behavior
of $\langle p_{{}_{T}}^2\rangle$ for Pb+Pb collisions.

In a hadron--nucleus collision, a parton from the projectile can
suffer soft quasielastic interactions as it crosses the nuclear
target on its way to the hard process.  We follow \cite{gg} and assume
that in each nucleon--nucleon, $NN$, subcollision there is a fixed
probability $\phi$ that the parton is affected.  In a hadron--nucleus
collision, the $\langle p_{{}_{T}}^{2}\rangle$ of the dimuon or $\psi$ is
then increased by:
\begin{equation} 
\Delta p_{{}_{T}}^{2} \equiv
\langle p_{{}_T}^2 \rangle -  \langle p_{{}_T}^2 \rangle_{{}_{NN}}
= \lambda^{2} ({\overline n}_A - 1),
\label{eq:pt1}
\end{equation}
where ${\overline n}_A$ is the number of $NN$ subcollisions that the
projectile suffers in the target and $\lambda^2\propto \phi$
determines the increment to the projectile parton's $p_{{}_T}^2$ from
each subcollision.  We subtract `1' to eliminate the hard subcollision
that produces the $\psi$ or dimuon from ${\overline n}_A$ and
introduce $\langle p_{{}_{T}}^{2}\rangle_{{}_{NN}}$ to account for the
$A$--independent contribution from that subcollision.  The impact
parameter averaged total number of subcollisions grows with the target
radius $R_{A} \approx 1.2\; A^{1/3}$ as ${\overline n}_A\approx
3\sigma_{{}_{NN}}\rho_0 R_{A}/2 \approx 0.77\; A^{1/3}$, where
$\rho_0$ is the average nuclear density and $\sigma_{{}_{NN}}\approx
32$~mb is the inelastic $NN$ cross section.  

In ref.~\cite{gg}, transverse momentum distributions measured in
hadron--nucleus collisions by CERN NA10 and NA3 \cite{pa} were used to
fix $\langle p_{{}_T}^2\rangle_{{}_{NN}}$ and $\lambda$ for Drell--Yan
and $\psi$ production.  In fig.~2, we compare $\Delta p_{{}_{T}}^{2}$
from eq.~(\ref{eq:pt1}) with more recent data from a larger number of
nuclear targets presented by FNAL E772 \cite{pat}.  The random walk
behavior, $\Delta p_{{}_{T}}^{2} \propto A^{1/3}$, is evident.  This
approximation agrees roughly with calculations using realistic nuclear
densities as in ref.~\cite{strik}.  In this work we take
$\lambda_{\mu^+\mu^-}\approx 0.18\pm 0.01$~GeV and
$\lambda_\psi\approx 0.36\pm 0.03$~GeV for Drell--Yan and $\psi$
production.  These values are chosen to agree with E772 and NA3 data,
respectively.  They are compatible with the idea that initial--state
scattering is a soft process.  We will use NA38 S+U data to obtain
$\langle p_{{}_T}^2\rangle_{{}_{NN}}$ below.

Observe that our Drell--Yan value is somewhat smaller than the
$0.24\pm 0.05$~GeV value used in ref.~\cite{gg}.  This spread perhaps
reflects the systematic uncertainty (not included in the fit uncertainty)
in comparing experiments with different kinematic coverages.  E772 did
not report $\Delta p_{{}_{T}}^2$ for $\psi$, but did give results for
$\Upsilon$ production, a process that likely has similar initial--state
interactions.  A fit to that data implies $\lambda_\Upsilon \approx
0.46$~GeV, which is larger than our $\lambda_\psi = 0.36$~GeV.  The
values that we have taken above are the smaller of the choices.  
\begin{figure}
\vskip -1.0in
\epsfxsize=3.3in
\centerline{\epsffile{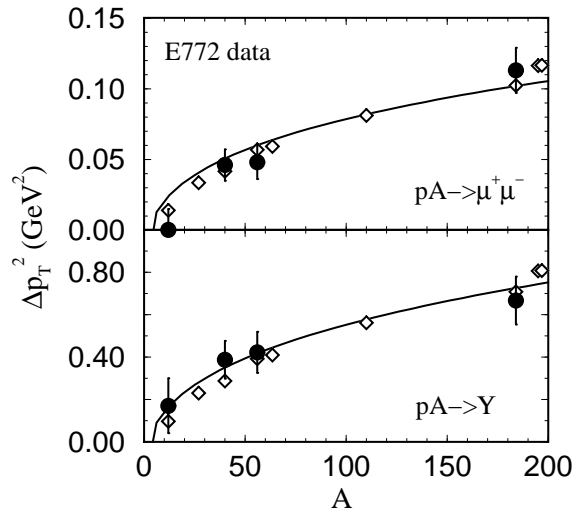}}
\vskip -0.8in
\caption[]{Calculated $\Delta p_{{}_{T}}^{2}$ from eq.~(\ref{eq:pt1})
for ${\overline n}_{A}\approx 0.77\; A^{1/3}$ compared to FNAL E772 data
on the Drell--Yan process and $\Upsilon$ production \cite{pat}.
Diamonds are calculated using (\ref{eq:nua}) and JVV nuclear densities
from \cite{jvv}.}
\end{figure}

We remark on the interpretation of the parameters in
eq.~(\ref{eq:pt1}). In ref.~\cite{gg}, the value of the ratio
$(\lambda_\psi/\lambda_{\mu^+\mu^-})^2\sim 2.3$ had been attributed to
the $9/4$ enhancement of the cross section for gluon--gluon relative
to quark--gluon forward scattering.  The initial--state partons are
predominantly gluons for $\psi$ production and quarks and antiquarks
for the Drell--Yan process, because the primary production mechanisms
for $\psi$ and dilepton production are gluon fusion and
quark--antiquark annihilation, respectively.  In contrast, the
parameters that we now use imply a ratio
$(\lambda_\psi/\lambda_{\mu^+\mu^-})^2\sim 3.9$ that is $\sim 70{\%}$
larger than the earlier estimate, while E772 values alone suggest
$(\lambda_\Upsilon/\lambda_{\mu^+\mu^-})^2\sim 6.8$.  The perturbative
estimate $(\lambda_\psi/\lambda_{\mu^+\mu^-})^2\sim 9/4$ \cite{pt,sg}
may not be correct in detail, since $\lambda^2$ involves small
momentum transfers.  Nevertheless, this disagreement is quite large.

If taken literally, a larger ratio can suggest that the $\psi$ also
undergoes final--state elastic scattering as it escapes the nucleus.
During its escape, the nascent $\psi$ is not a fully formed hadron
but, rather, a $c\overline{c}$ pair.  This pair can be in a color
octet state and, therefore, can scatter essentially as a gluon does.
Fortunately, this final--state octet scattering also follows
eq.~(\ref{eq:pt1}), so that we can describe this effect in conjunction
with initial--state gluon scattering by replacing $\lambda^2$ with
$2\lambda^2$.  Together with the $9/4$ color factor, this factor can
roughly account for the ratio of $\lambda^{2}$ in $\psi$ and
$\mu^{+}\mu^{-}$.  However, this observation is speculative.
Therefore, we will not include the factor of two explicitly, but
merely stress the possibility that the empirical $\lambda$ for
charmonium production receives contributions from both initial-- and
final--state octet scattering.
\begin{figure} 
\vskip -1.2in
\epsfxsize=3.5in
\centerline{\epsffile{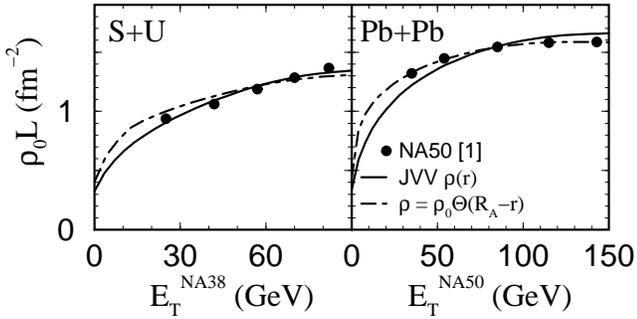}}
\vskip -1.5in
\caption[]{NA50 $L(E_T)$ [1] (points) compared to calculations
for the JVV nuclear densities \cite{jvv} (solid) and for
a sharp--surface approximation (dot--dashed).}
\end{figure}

In a nucleus--nucleus collision, both projectile and target partons
scatter.  We then write
\begin{equation}
\Delta p_{{}_{T}}^{2} \equiv
\langle p_{{}_{T}}^2 \rangle - \langle p_{{}_T}^2 \rangle_{{}_{NN}}
= \lambda^{2} ({\overline n}_A+{\overline n}_B - 2).
\label{eq:pt2}
\end{equation}
The NA50 path length is
\begin{equation}
L \equiv ({\overline n}_{A}+{\overline n}_{B})/2\sigma_{{}_{NN}}\rho_0
\equiv {\overline n}/2\sigma_{{}_{NN}}\rho_0.
\label{eq:let}
\end{equation}
The relation between ${\overline n}$ and the impact parameter $\vec{b}$
depends on the collision geometry.  A nucleon at a transverse 
position $\vec{s}$ in the projectile can undergo 
\begin{equation}
n_A = \sigma_{{}_{NN}}\int_{-\infty}^z\! dz^\prime\; \rho_{A}(\vec{s},
z^\prime),
\label{eq:nua}
\end{equation}
subcollisions prior to its hard interaction at longitudinal position
$z$, where $\rho_A$ is the projectile density.  The expression for the
number of subcollisions suffered by a nucleon at a transverse position
$\vec{b}-\vec{s}$ in the target is similar.  Since the hard process
occurs with a probability density $\propto
\rho_A(\vec{s},z)\rho_B(\vec{b}-\vec{s},z^\prime)$, the average number
of subcollisions for a given $b$ is
\begin{eqnarray}
{\overline n}(b) =  {{\sigma_{{}_{NN}}}\over {T_{AB}}}
\int\! d^2s\; T_A(s)&T_B&(|\vec{b}-\vec{s}|)[T_{A}(s)\nonumber\\
+ &T_{B}&(|\vec{b}-\vec{s}|)],
\label{eq:nu}
\end{eqnarray}
where $T_{A,\;B} = \int_{-\infty}^{\infty}\! dz^\prime\; \rho_{A,\; B}$
are the nuclear thickness functions and 
$T_{AB} = \int\! d^2s\; T_A(s)T_B(|\vec{b}-\vec{s}|)$.
An alternative derivation of eq.~(\ref{eq:nu}) in the spirit of ref.~\cite{gh}
is given in ref.~\cite{sg}.
\begin{figure} 
\vskip -1.2in
\epsfxsize=3.5in
\centerline{\epsffile{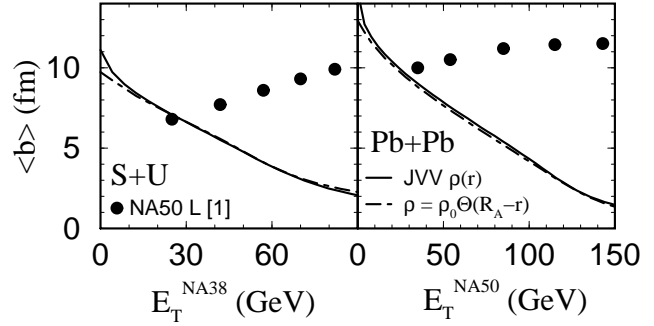}}
\vskip -1.5in
\caption[]{Average impact parameter as a function of $E_T$
for realistic nuclear densities (solid) and for
a sharp--surface approximation (dot--dashed). NA50 $L(E_T)$ [1] (points)
are included for comparison.}
\end{figure}

To obtain the $E_T$ dependence of ${\overline n}$ -- and therefore
$L(E_T)$ -- we fold eq.~(\ref{eq:nu}) with the probability $P(E_T,b)$ that
a collision at impact parameter $b$ produces transverse energy $E_T$.
This probability is related to the minimum--bias distribution by
\begin{equation}
\sigma_{\rm min}(E_{T}) = \int\! d^{2}b\; P(E_{T}, b).
\label{eq:min}
\end{equation}
The distribution $P(E_{T}, b)$ depends both on the collision geometry
and the calorimeter type.  We use the phenomenological $P(E_{T},b)$
relevant to the NA50 Pb+Pb collisions from ref.~\cite{gv2}; this
parametrization schematically incorporates fluctuations at fixed $b$
due to both collision dynamics and detector effects. 
The average number of collisions for a given $E_T$ is then
\begin{equation}
{\overline n}(E_T) = 
\sigma_{{}_{NN}}\langle T_A(s) + T_B(|\vec{b}-\vec{s}|)\rangle,
\label{eq:nuet}
\end{equation}
where we now average over the weighted product of the densities, 
\begin{equation}
\langle \ldots\rangle \equiv N^{-1}\int\! d^2bd^2s\, 
P(E_T,b)T_A(s)T_B(|\vec{b}-\vec{s}|)(\ldots),
\label{eq:avg}
\end{equation}
with $N = \int\! d^2b\, P(E_T,b) T_{AB}$ ({\it c.f.} eq.~(\ref{eq:nu})).

We now compute $L(E_T)$ using eqs.~(\ref{eq:let},\ref{eq:nuet}) and
(\ref{eq:avg}).  In fig.~3 we compare the NA50 $L(E_T)$ \cite{na50} to
the path length calculated using two assumptions for the nuclear
density profile.  Following ref.~\cite{gv2}, we use the realistic
three--parameter Woods--Saxon densities from deJager, deVries and
deVries (JVV) \cite{jvv}.  We compared this to a sharp--surface
approximation $\rho = \rho_0\Theta(R_A -r)$.  NA38~\cite{borhani}
obtained $L$ for S+U from $\psi$ $p_{{}_{T}}$ data using a
phenomenological procedure that is essentially equivalent to
eqs.~(\ref{eq:pt2}, \ref{eq:let}), while NA50 calculated $L$ assuming
the sharp--surface approximation~\cite{claudie}.  Consequently, we see
that the NA50 Pb+Pb values agree with our sharp--surface results, while
the NA38 S+U values are nearer to the realistic--density computations.
[The calculation of $L$ is not described in ref.~\cite{na50}, so the
agreement of our sharp--surface $L(E_T)$ with the NA50 Pb result is an
important check.]  Note that to compare to the NA50 calculations, we
plot $\rho_0L$ rather than $L$ because the authors of ref.~\cite{na50}
use a value $\rho_0 = 0.138$~fm$^{-3}$.

We see that $L(E_T)$ {\it saturates} in Pb+Pb collisions, as observed
in ref.~\cite{gv2}.  To see why saturation occurs in this system but
not in S+U, we compare the NA50 $L(E_T)$ \cite{na50} to the average
impact parameter $\langle b\rangle (E_T)$ in fig.~4.  We use the JVV
densities to compute $\langle b\rangle = \langle b
T_{AB}\rangle/\langle T_{AB}\rangle$.  [Note that NA50 reports similar
values of $\langle b\rangle (E_T)$ \cite{na50}.] For all but the
highest $E_T$ bin addressed by the S+U measurement, we see that
$\langle b\rangle$ is near $\sim R_{\rm S} = 3.6$~fm or larger.  In
this range, increasing $b$ dramatically reduces the collision volume
and, consequently, $L$.  In contrast, in Pb+Pb collisions $\langle
b\rangle < R_{\rm Pb} =$~6.6~fm for $E_T> 50$~GeV, so that $L$ does
not vary appreciably.
\begin{figure} 
\vskip -1.0in
\epsfxsize=3.3in
\rightline{\epsffile{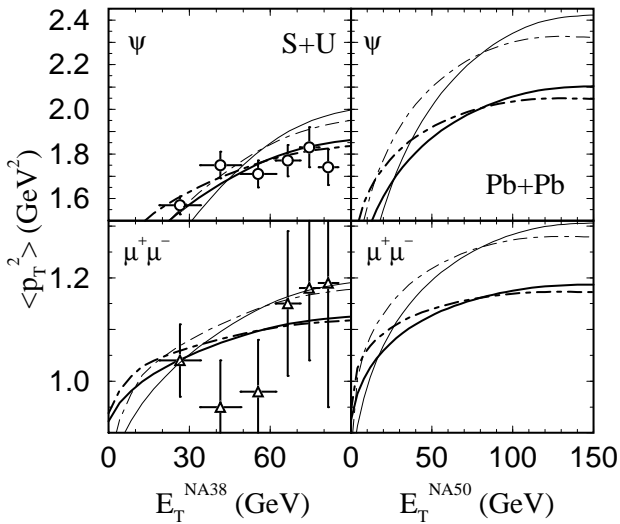}}
\vskip -0.6in
\caption[]{The calculated $E_T$ dependence of $\langle
p_{{}_{T}}^2\rangle$ for $\psi$ (upper plots) and Drell--Yan dimuons
(lower plots) in S+U and Pb+Pb collisions. The calculations are
performed for realistic nuclear densities (solid) and for the
sharp--surface approximation (dot--dashed).  NA38 S+U data is from
\cite{na38x}.  The thick curves are computed for our preferred values,
$\lambda_{\psi} = 0.36$~GeV and $\lambda_{\mu^+\mu^-} = 0.18$~GeV,
while the thin curves are for $\lambda_{\psi} = 0.46$~GeV and
$\lambda_{\mu^+\mu^-} = 0.24$~GeV.}
\end{figure}

We now use eqs.~(\ref{eq:pt2}, \ref{eq:nuet}) to compute $\Delta
p_{{}_{T}}^{2}$ for both the Drell--Yan process and $\psi$ production
as functions of $E_{T}$ in S+U and Pb+Pb collisions.  Our results are
shown in fig.~5.  Parton--scattering calculations agree with S+U data
from ref.~\cite{na38x}.  Observe that to compare to the NA38 S+U data,
we fit the data to extract $\langle p_{{}_{T}}^{2} \rangle_{{}_{NN}}
\approx 0.92$ and 1.07~GeV$^{2}$ for Drell--Yan pairs and $\psi$
respectively.  Note that the $\psi$ value is somewhat smaller than the
$1.23\pm 0.05$~GeV$^{2}$ reported by NA3 and used in ref.~\cite{gg};
this is likely another indication of systematic uncertainties.

To illustrate the spread in the Pb predictions implied by the
uncertainty in the parameters, we compute the thin curves in Fig.~5 by
taking $\lambda_\psi \approx \lambda_\Upsilon = 0.46$~GeV from E772
and $\lambda_{\mu^+\mu^-} = 0.24$~GeV from \cite{gg}.  With these
values, we obtain the best agreement with the NA38 data for $\langle
p_{{}_{T}}^{2} \rangle_{{}_{NN}} \approx 0.81$ and $0.6$~GeV$^2$ for
dimuons and $\psi$, respectively.  The modified $\psi$ results are
inconsistent with the S+U data. The modified dimuon results are
consistent, although the agreement is somewhat better for our
preferred value, $\lambda_{\mu^+\mu^-} = 0.18$~GeV.

The thick curves for Pb+Pb collisions in fig.~5 represent our
predictions.  We expect $\langle p_{{}_{T}}^2\rangle$ to increase by
$12.3{\%}$ for $\psi$ as $E_T$ increases from 50 to 150 GeV.  This
represents a flattening of $\langle p_{{}_{T}}^2\rangle(E_T)$ in
comparison to the S+U~$\rightarrow \psi+X$ data, which show an
$18.5{\%}$ increase as $E_T$ varies from 30 to 90 GeV.  The flattening
would be even more dramatic if the NA50 sharp--surface approximation
(the dot--dashed curve in fig.~5) were true --- $\langle
p_{{}_{T}}^2\rangle$ would increase by only $6.8{\%}$ for $\psi$ and
$2.7{\%}$ for dimuons.  While such small differences seem difficult to
resolve, agreement with S+U~$\rightarrow \psi+X$ data is better for
the realistic $L$ than for the sharp--surface result.  Furthermore,
NA38 \cite{borhani,claudie} used this $\psi$ data to obtain $L$ values
in agreement with our realistic calculation and quite distinct from
our sharp--surface result.  However, we stress that the most direct
extraction of $L$ comes not from $\psi$ but from Drell--Yan dimuons.
To check that the NA50 plot \cite{na50} is correct, {\it i.e.} to
decide between the open circles and filled circles in fig.~1,
Drell--Yan data in Pb+Pb would have to establish a $2.7{\%}$ increase
in $\langle p_{{}_{T}}^2\rangle$.  This requires dimuon data far more
precise than that for S+U collisions.

One can ask if elastic scattering of the $\psi$ by comovers can affect
its transverse momentum distribution.  As mentioned earlier, comover
absorption has only a marginal effect on the $p_{{}_{T}}$
distribution; its influence on $\langle p_{{}_{T}}^{2}\rangle$ is
negligible \cite{ggj,gg}.  However, elastic comover scattering can in
principle allow the $\psi$ to be pushed transversely, adding to its
$p_{{}_T}$.  The argument \cite{ggj,plasma} that elastic scattering
with hadronic comovers is negligible follows from photoproduction
data.  The measured elastic $\psi N$ cross section is $\sim 0.079 \pm
0.012$~mb \cite{emc}, accounting for only $\sim 2-4{\%}$ of the total
cross section.  The corresponding mean--free path for elastic
scattering with hadronic comovers greatly exceeds the estimated size
of the system, so that the $\psi$ will not follow the comover flow.
The general relation between scattering and flow is discussed in
ref.~\cite{prakash}.  One possible hole in this argument is that $\psi
\pi$ scattering in the comover gas is likely below the $D{\overline
D}$ dissociation threshold and, therefore, predominantly elastic.
Another possible hole is that comovers need not be hadrons.

The consistency of the NA38 S+U~$\rightarrow \psi+X$ data in fig.~5
with (\ref{eq:pt2}) supports our neglect of elastic comover
scattering.  Nevertheless, if Pb+Pb experiments find a $\langle
p_{{}_{T}}^{2}\rangle$ for $\psi$ that is larger than our prediction
{\it and} if other hadronic species show evidence of substantial
transverse flow, it will be necessary to introduce elastic scattering
into the comover scenario.  This is best done in the context of
cascade models.

In summary, we have predicted the nuclear enhancement of $\langle
p_{{}_{T}}^2\rangle$ in Pb+Pb collisions as a function of $E_T$ for
charmonia and Drell--Yan dimuons, assuming that this enhancement is
caused by quasielastic parton scattering.  Such scattering has been
included \cite{gg,pt,plasma} in hadronic models of $\psi$ suppression,
where it is essential for describing the $p_{{}_T}$ dependence of
$p$A, O+U and S+U data. We stress that the Drell--Yan process is
unaffected by final--state comover or plasma interactions, so that a
comparison of both $\psi$ and $\mu^+\mu^-$ calculations to data can
disentangle model uncertainties in $L(E_T)$ from new physics.  In
particular, the $L$ extracted from dimuon measurements can confirm or
disprove the threshold behavior claimed by NA50.

The parameters in the model were revised from an earlier work
\cite{gg} to describe the latest E772 $p$A~$\rightarrow \mu^+\mu^- +
X$ data.  Our revision implies a ratio
$(\lambda_\psi/\lambda_{\mu^+\mu^-})^2\sim 3.9$, larger than that
extracted earlier \cite{gg}, perhaps indicating that additional
final--state scattering of the octet $c\overline{c}$ occurs in the
charmonium case.  More precise $p$A~$\rightarrow \psi + X$
measurements are needed to explore this very interesting possibility.

We are grateful to Claudie Gerschel for discussing details of the
NA38/50 data and to Mark Strikman for his careful reading of the
manuscript.  We also thank Miklos Gyulassy, Raffaele Mattiello and
Bill Zajc for discussions.  This work was supported in part by the
U. S. Department of Energy under Contract Numbers DE-FG02-93ER40764
and DE-AC03-76SS0098.

\end{narrowtext}
\end{document}